\documentclass[11pt]{article}
\usepackage{amssymb}

\newcommand{\half}{{\scriptstyle{\frac{1}{2}}}}

\newcommand{\dl}{\delta^{(3)}({\bf r})}

\newcommand{\BE}{\begin{equation}}
\newcommand{\EE}{\end{equation}}
\newcommand{\BA}{\begin{eqnarray}}
\newcommand{\EA}{\end{eqnarray}}
\newcommand{\vol}{{\sf V}}
\newcommand{\num}{{\sf N}}

\begin{document}
\begin{titlepage}

\vspace*{1mm}
\begin{center}

            {\LARGE{\bf Quantum-hydrodynamical picture of the
             \\ massive  Higgs boson}}

\vspace*{14mm}
{\Large  M. Consoli and E. Costanzo }
\vspace*{4mm}\\
{\large
Istituto Nazionale di Fisica Nucleare, Sezione di Catania \\
and Dipartimento di Fisica, Citt\`a Universitaria \\
Via Santa Sofia 64, 95123 Catania, Italy}
\end{center}
\begin{center}
{\bf Abstract}
\end{center}
The phenomenon of spontaneous symmetry breaking admits a physical interpretation
in terms of the Bose-condensation process of elementary spinless quanta. 
In this picture, the broken-symmetry phase emerges as a real physical medium, 
endowed with a hierarchical pattern of scales, 
supporting two types of elementary excitations for ${\bf{k}} \to 0$: 
a massive energy branch $E_a({\bf{k}}) \to M_H$, 
corresponding  to the usual Higgs boson field, 
and a collective gap-less branch $E_b({\bf{k}}) \to 0$. This is similar
to the coexistence of phonons and rotons in
superfluid $^4$He that, in fact, is usually considered the condensed-matter
analog of the Higgs condensate. 
 After previous work dedicated to the properties 
of the gap-less, phonon branch, in this paper we use quantum hydrodynamics to
propose a physical 
interpretation of the massive branch. On the base of our results, 
$M_H$ coincides with the energy-gap for vortex formation and
a massive Higgs boson is like a roton in superfluid $^4$He. Within this
interpretation of the Higgs particle, 
there is no {\it naturalness} problem since
$M_H$ remains a naturally intermediate, fixed 
energy scale, even for an ultimate ultraviolet 
cutoff $\Lambda \to \infty$.
\vskip 35 pt
\end{titlepage}

\section{Introduction}

The idea of a spontaneously broken phase as a `condensate' is now widely
accepted. For instance, 
in the physically relevant case of the Standard Model,
the situation can be summarized saying \cite{thooft} that 
 "What we experience as 
empty space is nothing but the configuration of the Higgs field that has the
lowest possible energy. If we move from field jargon to particle jargon, this 
means that empty space is actually filled with Higgs particles. They have 
Bose condensed."  

Here, by `Bose condensation' one means the phenomenon
of spontaneous particle creation in the same quantum state
(${\bf{k}}=0$ in some reference frame) 
from the empty vacuum $|o\rangle$ of perturbation theory. In this way, 
the translation from `field jargon to particle jargon', 
amounts to establish a well defined functional relation 
(see ref.\cite{mech} and Sect.2)
$n=n(\phi^2)$ between the average density of scalar quanta, the `phions', 
and the average value of the scalar 
field $\langle \Phi \rangle=\phi$. 
Thus, Bose condensation is just a consequence of
the minimization condition of the 
 effective potential $V_{\rm eff}(\phi)$. This has absolute
minima at some values $\phi =\pm v \neq 0$ for which $n(v^2)=\bar{n}\neq 0$
\cite{mech}. 

Of course, in order this picture to be consistent, spontaneus symmetry 
breaking should occur 
for physical (i.e. real and non-negative) values of the phion mass $m_\Phi$.  
In other words, Bose condensation requires the phase transition in
(cutoff) $\lambda\Phi^4$ theories  to be
{\it first order}. While in the presence of gauge bosons 
this can be shown perturbatively \cite{cw}, 
the use of perturbation theory in a pure $\lambda\Phi^4$ theory 
leads to contradictory results between even and odd orders \cite{trivpert}.  
Therefore, one has
to analyze the effective potential beyond perturbation theory. 
By relying on the assumed exact `triviality' property of the theory in 
3+1 space-time dimensions \cite{book}, one is driven to consider
the class of `triviality compatible' approximations 
\cite{zeit,plb,mech} 
to the effective potential, say
$V_{\rm eff}(\phi)=V_{\rm triv}(\phi)$. This includes the 
one-loop potential, the gaussian and 
post-gaussian \cite{ritschel} calculations of the
Cornwall-Jackiw-Tomboulis \cite{cjt} effective potential for composite
operators, i.e. all approximations 
where the fluctuation field is governed by an effective
quadratic Hamiltonian. In all such cases, 
the lowest energy state of the 
massless theory at $m_\Phi=0$ corresponds to a broken-symmetry phase as
first shown in ref.\cite{cian} for the gaussian approximation. 
Therefore the phase transition, occurring earlier for small positive
values of $m_\Phi$, is first order.
 However, there is a subtlety since it corresponds to an
{\it infinitesimally weak} first-order phase transition. 
In fact, it is first order for any finite ultraviolet cutoff $\Lambda$ 
but becomes asymptotically
second order \cite{notecw}
in the continuum limit $\Lambda \to \infty$. 

In any case, for all finite values of $\Lambda$, 
a {\it particle-gas} picture
of the underlying scalar system is possible. This 
can be formulated in terms of two basic quantities: the equilibrium 
phion density $\bar{n}$ and the phion-phion scattering length $a$
\BE
\label{aa}
a = \frac{\lambda}{8 \pi m_\Phi}  
\EE
defined, in the limit of zero-momentum scattering, from the dimensionless
scalar self-coupling $\lambda$ and the phion mass. 

As shown in ref.\cite{mech},
these two quantities combine to produce all relevant length scales of the
system
\BE
\label{scales}
a \ll {{1}\over{ \sqrt{ \bar{n}a} }} \ll {{1}\over{ \bar{n}a^2}}
\EE
that decouple for an infinitely dilute system where 
$\bar{n}a^3 \to 0$. In this situation, which corresponds to  approaching
the continuum limit of quantum field theory \cite{mech},  
one discovers an unexpected result: the 
{\it hierarchical} nature of the scalar condensate.

At the same time, a particle-gas picture of the broken-symmetry phase
raises several questions. For instance, in condensed 
media the properties of the system over vastly different
scales are described by different branches of the energy spectrum. Are there
similar transitions in the scalar condensate ? Also, what 
about the coexistence of
{\it exact} Lorentz covariance and vacuum condensation in {\it effective}
quantum field theories ? 
Can the violations of locality, at the energy scale fixed by the
ultraviolet cutoff, induce
non-Lorentz-covariant modifications of the
{\it infrared} energy spectrum 
that depends on the vacuum structure \cite{salehi} ?

To indicate this type of infrared-ultraviolet connection, 
originating 
from vacuum condensation in effective quantum field theories, 
Volovik \cite{volo1} has introduced a very appropriate name: 
reentrant violations of special relativity in the low-energy corner. 
These occur in a small shell of three-momenta, 
say $|{\bf{k}}| < \delta$, 
that only vanishes in the strict
local limit where $\Lambda \to \infty$ and an exact 
Lorentz-covariant energy spectrum
is re-obtained in the whole range of momenta. 

By denoting $M_H$ as the typical energy scale associated with the 
Lorentz-covariant part of the energy spectrum, the `reentrant' nature of
the violating effects means that
${\cal O}({{\delta}\over{M_H}})$ vacuum-dependent corrections are
equivalent to 
 ${\cal O}({{M_H}\over{\Lambda}})$ effects (see below).
  The $1/\Lambda$ terms, that have always been
neglected when discussing \cite{nielsen} how
Lorentz covariance emerges in effective theories when removing
the ultraviolet cutoff, although infinitesimally small, can play an 
important role over distances larger than $\Lambda/M^2_H$, i.e.
infinitely larger than the typical elementary particle scale $\xi_H=1/M_H$. 

As discussed in ref.\cite{weak,hierarchy,drift}, 
the basic ingredient to understand the nature of the `reentrant' 
effects in the scalar condensate consists in a purely
quantum-field-theoretical result: the {\it two-valued} nature of 
the connected zero-four-momentum propagator $G^{-1}(k=0)$ in the broken phase
\cite{legendre,pmu}. 
In fact, besides the well known massive solution
$G^{-1}_a(k=0)= M^2_H$, one also finds
$G^{-1}_b(k=0)= 0$. 

The b-type of solution
corresponds to processes where assorbing (or emitting)
a very small 3-momentum ${\bf{k}} \to 0$
does not cost any finite energy. This situation is well
known in a condensed medium, where a small 
3-momentum can be coherently distributed 
among a large number of elementary constituents, and corresponds to 
the hydrodynamical regime of density fluctuations whose
wavelengths $2\pi/|{\bf{k}}|$ are {\it larger} 
than $r_{\rm mfp}$, 
the mean free path for the elementary constituents. 

In this sense, the situation is very similar to
superfluid $^4$He, 
where the observed energy spectrum is due to the peculiar
transition from the `phonon branch' to the `roton branch' at a momentum scale 
$|{\bf{k}}_o|$ where
${E}_{\rm phonon}({\bf{k}}_o) \sim 
{E}_{\rm roton}({\bf{k}}_o)$. 
The analog for the scalar condensate amounts to
an energy spectrum with the following limiting behaviours :

~~~i) ${E}({\bf{k}}) \to {E}_b({\bf{k}}) \sim c_s |{\bf{k}}|$   
       ~~~~~~~~~~~~~~~~~~~~ for ${\bf{k}}\to 0 $

~~~ii) ${E}({\bf{k}}) \to {E}_a({\bf{k}}) \sim M_H+ {{ {\bf{k}}^2 }\over{2 M_H}}$
       ~~~~~~~~~~ 
for $|{\bf{k}}| \gtrsim \delta $

where the 
characteristic momentum scale $\delta \ll M_H$, at which
$E_a(\delta)\sim E_b(\delta)$, marks the transition from collective to 
single-particle excitations. This occurs for
\BE
\label{dd}
\delta \sim 1/r_{\rm mfp} 
\EE
 where \cite{kine,seminar}
\BE
\label{mfp}
r_{\rm mfp} \sim {{1}\over{ \bar{n} a^2}}
\EE
is the phion mean free path, 
for a given value of the phion density $n=\bar{n}$ and a
given value of the phion-phion scattering length $a$. In terms of the 
same quantities, one also finds (see \cite{mech} and Sect.2)
\BE
               M^2_H \sim \bar{n} a
\EE
giving the anticipated trend of the dimensionless ratios ($\Lambda \sim 1/a$)
\BE
\label{golden}
{{\delta}\over{M_H}} \sim {{M_H}\over{\Lambda}} \sim \sqrt{ \bar{n}a^3} \to 0
\EE
in the continuum limit where $a \to 0$ and the mass scale $\bar{n}a$ is
held fixed \cite{compact}.

Now, deducing the detailed form of the energy spectrum 
that interpolates between 
${E}_a({\bf{k}})$ and ${E}_b({\bf{k}})$ is a formidable task. 
 To have an idea, the
same problem in superfluid $^4$He, after
more than fifty years and despite the efforts of many theorists, notably
Landau and Feynman, has not been solved in a satisfactory way. 
Therefore, 
by taking into account the above remark, one can simply 
approximate \cite{weak,hierarchy,drift} the expansion of the scalar field in 
the broken phase by two separate branches 
as (phys=`physical') \cite{physical}
\BE
\label{phime2}
\Phi_{\rm phys}(x) = v_R + {h}(x) + {H}(x)
\EE 
with 
\BE
\label{hh}
{h}(x)=
\sum_ { | {\bf {k}}| < \delta }  
\frac{1} { \sqrt{2 \vol {E}_k } } 
\left[  \tilde{h}_{\bf k}    {\rm e}^ { i ({\bf k}.{\bf x} -{E}_k t) } + 
(\tilde{h}_{\bf k})^{\dagger} {\rm e}^{-i ({\bf k}.{\bf x} -{E}_k t)} 
\right]
\EE
and
\BE
\label{H2}
{H}(x)=
\sum_{ |{\bf {k}}| > \delta }  
\frac{1} { \sqrt{2 \vol {E}_k } } 
\left[  \tilde{H}_{\bf k}    {\rm e}^ { i ({\bf k}.{\bf x} -{E}_k t) } + 
(\tilde{H}_{\bf k})^{\dagger} {\rm e}^{-i ({\bf k}.{\bf x} -{E}_k t)} 
\right]
\EE
where $\vol$ is the quantization volume and 
${E}_k=c_s|{\bf{k}}|$ for $|{\bf{k}}| < \delta$ while
${E}_k=\sqrt{{\bf{k}}^2 + M^2_H}$ for $|{\bf{k}}| > \delta$. Also, 
$c_s \delta \sim M_H$. 

Eqs.(\ref{phime2})-(\ref{H2}) replace
the more conventional relations
\BE
\label{conve1}
\Phi_{\rm phys}(x)= v_R + H(x)
\EE 
where
\BE
\label{conve2}
{H}(x)=
\sum_{ {\bf {k}} }
\frac{1} { \sqrt{2 \vol {E}_k } } 
\left[  \tilde{H}_{\bf k}    {\rm e}^ { i ({\bf k}.{\bf x} -{E}_k t) } + 
(\tilde{H}_{\bf k})^{\dagger} {\rm e}^{-i ({\bf k}.{\bf x} -{E}_k t)} 
\right]
\EE
with ${E}_k=\sqrt{{\bf{k}}^2 + M^2_H}$. 
Eqs.(\ref{conve1}) and (\ref{conve2})
are reobtained in the limit 
${{\delta}\over{M_H}} \sim {{M_H}\over{\Lambda}} \to 0$ 
where the wavelengths associated to
$h(x)$ become infinitely large in units of the physical scale set by 
$\xi_H=1/M_H$.  In this limit, where for any finite value of
${\bf{k}}$ the broken phase has
only massive excitations, one recovers an exactly Lorentz-covariant theory.

Now, as anticipated, 
the interpretation of the gap-less branch in terms of density fluctuations
places no particular problem \cite{weak,hierarchy,drift}. In fact, 
"Any quantum liquid consisting of particles with integral 
spin (such as the liquid isotope $^4$He) must certainly have a spectrum of
this type...In a quantum Bose liquid, elementary excitations with small 
momenta ${\bf{k}}$  (wavelengths large compared with distances between atoms) 
correspond to ordinary hydrodynamic sound waves, i.e. are phonons. This 
means that the energy of such quasi-particles is a linear function of their
momentum" \cite{pita}. In this sense, a superfluid 
vacuum provides for ${\bf{k}} \to 0$ a universal picture.
This result does not depend
on the details of the short-distance interaction and even on the nature
of the elementary constituents. 
For instance, the same coarse-grained description is found in 
superfluid fermionic vacua \cite{volo2} that, as compared to the Higgs 
vacuum, bear the same relation of superfluid $^3$He to superfluid $^4$He.

On the other hand, a full analogy with $^4$He would also require to 
establish the interpretation of the massive branch 
for $|{\bf{k}}| \gtrsim \delta $ as a `roton', i.e. in terms of
a suitable vortical motion in the superfluid, and this
 is by no means obvious. For instance, Landau's roton
spectrum 
\BE
\label{roton}
{E}_{\rm roton}({\bf{k}})= \Delta +  {{ {\bf{k}}^2 }\over{2\mu}}
\EE
depends on two parameters $\Delta$ and $\mu$ that, 
in superfluid $^4$He, are vastly different. In fact, 
(in units $\hbar=c=1$, where $\Delta$ and $\mu$ have the same physical 
dimensions) one finds $\Delta\sim 7\cdot 10^{-4}$ eV and
$\mu\sim 6\cdot 10^{8}$ eV while
${E}_a({\bf{k}}) = \sqrt{ {\bf{k}}^2 + M^2_H}$ depends on a single mass
parameter $M_H$. 
Under which conditions can a superfluid medium exhibit rotons with
$\Delta=\mu$ ? Moreover, even if $\Delta=\mu$, does this value agree with
the Higgs mass parameter $M_H$ obtained in quantum field theory ?

The answer to this type of questions can only be obtained by combining 
a field-theoretical description of the condensation phenomenon with the basic
ingredients of quantum hydrodynamics. This analysis represents the main
content of this paper and will be presented in the following. 
In Sect.2, we shall first review the formalism of ref.\cite{mech} with
the hierarchical pattern of scales that is established in 
the scalar condensate. Further, in Sect.3, we shall use
the formalism of quantum hydrodynamics and discuss the interpretation
of the massive branch as a roton. Finally, Sect.4 will contain a summary
together with other possible consequences of our approach.

\section{The Higgs condensate and a hierarchy of scales}

We shall now first
resume the main results of ref.\cite{mech} in the case of
a one-component $\lambda \Phi^4$ theory, a system 
 where the condensing quanta are just neutral spinless 
particles, the `phions'. 

One starts by quantizing 
the scalar field $\Phi({\bf{x}})$ in terms of 
$a_{\bf k}$, $a^{\dagger}_{\bf k}$, the 
annihilation and creation operators for the elementary phions
whose `empty' vacuum state $|o\rangle$ is defined through
$a_{\bf k}|o\rangle=\langle o|a^{\dagger}_{\bf k}$=0. 

The phion system is assumed to be contained within a finite box of volume $\vol$ 
with periodic boundary conditions.  There is then a discrete set of allowed 
modes ${\bf k}$.  In the end one takes the infinite-volume limit and the 
summation over allowed modes goes over to an integration: 
$\sum_{\bf k} \to \vol \int d^3k/(2 \pi)^3$.  In this way, the scalar field
is expressed as
\BE
\label{phime}
\Phi({\bf x}) = 
\sum_{\bf k} \frac{1}{\sqrt{2 \vol E_k}} 
\left[ a_{\bf k} {\rm e}^{i {\bf k}.{\bf x}} + 
a^{\dagger}_{\bf k} {\rm e}^{-i {\bf k}.{\bf x}} 
\right] ,
\EE
where $E_k=\sqrt{{\bf k}^2 + m^2_\Phi}$, $m_\Phi$ 
being the physical, renormalized phion mass.

Bose condensation means that in the ground state
 there is an average number $\num_0$ of phions in the ${\bf k}=0$ 
mode, where $\num_0$ is a finite fraction of the 
total average number $\num$
\BE
         \num=\langle 
\sum_{\bf{k}} a^{\dagger}_{ {\bf{k}} } a_{ {\bf{k}} } \rangle
\EE
At zero temperature, if the gas is dilute, almost all the particles are in 
the condensate; $\num_0{ (T=0)} \sim \num$.  In fact, the fraction 
which is not in the condensate (`depletion')
\BE
\label{deple}
            D= 1- {{\num_0}\over{\num}} ={\cal O}(\sqrt{n a^3})
\EE
is a phase-space effect
that becomes negligible for a very dilute system \cite{bose} where
\BE
\label{eepsilon}
                 \epsilon = \sqrt {n a^3} \ll 1
\EE
In Eqs.(\ref{deple}) and (\ref{eepsilon}) 
 we have introduced the phion density
\BE
\label{density}
                        n= {{\num}\over{\vol}}
\EE
and 
the phion-phion scattering length Eq.(\ref{aa}).

Therefore, for a very dilute system, where, to a first approximation, one 
neglects the residual operator part of $a_{ {\bf{k}}=0}$, one gets 
$a^{\dagger}_{ {\bf{k}}=0} a_{ {\bf{k}}=0} \sim \num$ and so, 
 $a_{ {\bf{k}}=0}$ can be identified with
the c-number $\sqrt{\num}$ (up to a phase factor).  In this way
\BE
\label{phn}
\phi = \langle \Phi \rangle = \frac{1}{\sqrt{2 \vol m_\Phi}} 
\langle (a^{\dagger}_{ {\bf{k}}=0}+ a_{ {\bf{k}}=0}) \rangle \sim  
\sqrt{\frac{2\num}{\vol m_\Phi}}.
\EE
or 
\BE
\label{neq}
n(\phi^2) \sim \half m_\Phi \phi^2,
\EE
With this identification, setting 
$a^{\dagger}_{ {\bf{k}}=0}= a_{{\bf{k}}=0}=\sqrt{\num}$ is equivalent to 
shifting the quantum field $\Phi$ by a constant term $\phi$. 
Further, using Eq.(\ref{neq}), 
the energy density ${\cal E}={\cal E}(n)$
 can be translated into the
effective potential 
\BE
V_{\rm eff}(\phi)={\cal E}(n)
\EE
Now, by exploring the limit $m_\Phi \to 0$ in the class of 
`triviality compatible'
approximations to the effective potential 
\cite{zeit,plb,mech}
one discovers non-trivial absolute minima
$\phi=\pm v \neq 0$ of
$V_{\rm eff}(\phi)=V_{\rm triv}(\phi)$ and
 Bose condensation 
with an average density $\bar{n}=n(v^2)$. 

The basic relations of the broken phase are 
\BE
\label{mh}
                     M^2_H \sim \lambda v^2 \sim \bar{n}a 
\EE
and 
\BE
\label{mc}
                m^2_\Phi \sim \lambda^2 v^2 \sim \epsilon^2 \bar{n}a 
\EE
The key-ingredient to understand why 
the phase transition occurs for a still positive value of $m_\Phi$, 
consists in the observation \cite{mech} that the 
phion-phion interaction is not always repulsive. Besides the contact 
$+\lambda\dl$ potential there is an induced attraction 
$-\lambda^2 {{e^{-2m_\Phi r}}\over{r^3}}$ from ultraviolet-finite parts of
higher order graphs (see also ref.\cite{ferrer}) that, differently from 
the usual ultraviolet divergences, cannot be
re-absorbed into a standard redefinition of 
the tree-level coupling. For small values of $\lambda$ and
for sufficiently small values of $m_\Phi$
the corresponding graphs,  
 when taken into account consistently in the 
effective potential, can compensate for the effects of
both the short range repulsion and of
the non-zero phion mass. In this 
situation, the perturbative empty vacuum state $|o\rangle$, although
locally stable, is not globally stable
and the lowest energy state becomes a 
state with a non-zero density of phions that
are Bose condensed in the zero-momentum state. 

We emphasize that this weakly first-order scenario of symmetry breaking 
is discovered in a {\it class} of approximations to the effective 
potential, just those that are consistent with the assumed exact `triviality' 
property of the theory in 3+1 space-time dimensions \cite{book}.
In any case, it can be objectively tested against
the standard picture
based on a second-order phase transition. To this end one can run 
numerical simulations near
the phase transition region and compare the predictions of 
refs.\cite{zeit,plb} with the conventional existing two-loop or 
renormalization-group-improved forms of the effective potential. 
When this is done, the quality of the fits to the
existing lattice data  \cite{agodi,fiore} 
favours unambiguosly the first-order scenario of refs.\cite{zeit,plb,mech}. 

Let us now consider the range of momenta associated with the two different
branches  of the energy spectrum. In condensed matter, 
the transition between acoustic branch and single-particle
spectrum corresponds to their matching at 
a momentum scale set by the inverse 
mean free path for the elementary constituents. As anticipated in the 
Introduction, 
in the scalar condensate, the matching condition corresponds to 
a momentum scale $\delta\ll M_H$ where  $E_a(\delta) \sim E_b(\delta)$ or
\BE
\label{match}
               c_s \delta \sim \sqrt {\delta^2 + M^2_H} \sim M_H + 
{{\delta^2}\over{2M_H}}
\EE
with
${\delta} \sim {{1}\over{ r_{\rm mfp} }}$, 
$r_{\rm mfp}$ being
 the phion mean free path for $n=\bar{n}$ 
in Eq.(\ref{mfp}). 

Now, the scattering length $a$ 
can be used to define a far ultraviolet scale
\BE
\Lambda \equiv 1/a
\EE
up to which phions can be treated as `hard spheres'. Using Eqs.(\ref{dd}), 
(\ref{mfp}), (\ref{mh}), and (\ref{match}), this yields
\BE
\label{tt}
   t= {{\Lambda}\over{M_H}} \sim \sqrt{  {{1}\over{ \bar{n} a^3}} } 
\EE
and 
\BE
\label{epsilon}
{{1}\over{c_s}}\sim {{\delta}\over{M_H}} \sim 
\sqrt { \bar{n}a^3} 
\EE
Therefore, in the continuum limit where $t \to \infty$ 
one gets an infinitely dilute
Higgs condensate where 
$\epsilon=\sqrt{\bar{n}a^3} \to 0$ and
the hierarchy of scales 
\BE
\delta \ll M_H \ll  \Lambda
\EE
 related as in Eq.(\ref{golden}). 

Finally, by using Eq.(\ref{aa}), the condition for 
spontaneous symmetry breaking 
Eq.(\ref{mc})
can also be expressed as
\BE
\label{mrmfp}
 m_\Phi \sim \bar{n}a^2 \sim 
{{1}\over{r_{\rm mfp} }}
\EE
so that the phion mean free path in the condensate
is of the same order as the phion Compton
wavelength $1/m_\Phi$.

Notice that the order of magnitude of the
sound velocity
\BE
\label{cs}
       c_s \sim {{M_H}\over{\delta}} \sim {{\Lambda}\over{M_H}} \sim 
{{1}\over{\epsilon}}
\EE
is much larger than unity (in units of the light velocity $c=1$). 
Actually, $c_s$ {\it must} diverge 
in the continuum limit where Lorentz-covariance becomes exact \cite{weak}.
 In this limit,
where the vacuum acquires an infinite rigidity, the condensate becomes
incompressible and the 
massive branch of the energy spectrum remains valid
down to ${\bf{k}}=0$. 

The presence of superluminal sound in the scalar condensate
 has different motivations. First of all we observe that, on a general ground, 
"..it is an open question whether 
${{c_s}\over{c}}$ remains less than unity when nonelectromagnetic forces are
taken into account" \cite{weisound}. For this reason, 
several authors \cite{poly,bludman1,keister} have considered
the possibility of media whose long-wavelength 
compressional modes for ${\bf{k}} \to 0$
have phase and group velocity 
${{{E}}\over{ |{\bf{k}}| }}=
{{d{E}}\over{ d|{\bf{k}}| }}=c_s>c$. 

This possibility depends on the approximate nature of locality 
in cutoff-dependent quantum field theories where the elementary
quanta are treated as `hard spheres'. In this case, in fact, 
a hard-sphere radius is known \cite{born} 
to imply a superluminal propagation
within the sphere boundary. Now, in the perturbative empty vacuum state 
(with no
condensed quanta) such superluminal propagation is restricted to very short
wavelengths, smaller than the inverse ultraviolet cutoff. However, in the
condensed vacuum, the hard spheres can `touch' each other so that the 
actual propagation of 
density fluctuations in a hard-sphere system
might take place at a superluminal speed. This intuitive idea is at the
base of the `macroscopic' violations of locality discussed in 
ref.\cite{poly}
(and of the `reentrant' violations of special relativity in the low-energy
corner \cite{volo1} mentioned above). 

In some case, superluminal sound is known to arise  \cite{bludman1}
when a large negative bare mass and a large positive self-energy 
combine to produce a very small physical mass, just 
the situation expected for the quanta of 
the scalar condensate. In this way, the physical
origin of the superluminal sound is traced back to the 
asymmetric role of mass renormalization: it subtracts out self-interaction
energy without altering the tree-level
interparticle interactions that contribute
to the pressure. 

On the other hand, 
following refs.\cite{seminar,weak}, superluminal sound 
is also consistent with the
equation of state of a perfect fluid whose energy density has a minimum 
at some given value of the particle density, as it happens in the scalar
condensate. Just for this reason, near the minimum, 
long-wavelength density fluctuations represent nearly instantaneous effects
that can propagate at arbitrarily large speed. In this sense, the scalar
condensate rensembles an elastic medium near the inconpressibility limit
where the Poisson ratio $\nu \to 1/2$ and the propagation speed of the 
longitudinal waves of dilatation diverges in units of the propagation speed
of the transverse waves of distortion \cite{hierarchy}. 

Summarizing the previous results, we find that
in the local limit of the theory, where $\Lambda/M_H \to \infty$, 
one also finds $\delta/M_H \to 0$ so that
the energy spectrum $E({\bf{k}})$ reduces to 
$E_a({\bf{k}})=\sqrt{ {\bf{k}}^2 + M^2_H}$ 
in the whole range of $|{\bf{k}}|$.
In the cutoff theory, however, 
one should expect infinitesimal deviations
in an infinitesimal region of three-momenta. 
For instance, assuming
$\Lambda=10^{19}$ GeV and $M_H=250$ GeV, a scale 
$\delta\sim {{M^2_H}\over{\Lambda}}\sim 10^{-5}$ eV, for which
${{\delta}\over{M_H}}\sim 4\cdot10^{-17} $, 
 might well represent the physical 
realization of a formally infinitesimal quantity. If this were
the right order of magnitude, 
the collective density fluctuations of the Higgs vacuum described by
$E_b({\bf{k}})$
have wavelengths $> {{2\pi}\over{\delta}}$ 
thus ranging from about a centimeter up to infinity 
\cite{weak,hierarchy,drift}.

On the other hand, 
for $|{\bf{k}}| \gtrsim \delta$, the excitation spectrum describes 
single-particle states of mass $M_H\sim \sqrt{\bar{n}a}$. In the following
section,
we shall show that these states can be interpreted as elementary excitations
associated with a vortical motion.

\section{The `roton' picture of the massive branch}

In his theory, Landau suggested that in a superfluid medium 
there must be elementary vortex
excitations, the `rotons', whose energy has the form in Eq.(\ref{roton}).
In his original papers \cite{hydrolandau}, Landau did not work out
an explicit
derivation of Eq.(\ref{roton}). This was, however, deduced subsequently 
by Ziman \cite{ziman} whose formalism we shall briefly resume in the
following (for the convenience of the reader we shall adopt in this section 
the same notations of ref.\cite{ziman}). 

Ziman's starting point is the form of the Hamiltonian density of a fluid
\BE
                   H= {{1}\over{2}} \rho {\bf{u}}^2 +\rho W(\rho)
\EE
where $\rho$ is the mass density and $W$ the internal energy
whose mimimum is obtained for $\rho \equiv \bar{\rho}$. 
The fluid velocity
\BE
\label{clebsch}
    {\bf{u}}=-\nabla \varphi -{{i}\over{2\rho }}
(\Psi^* \nabla \Psi - \Psi \nabla \Psi^*)
\EE
is expressed 
in terms of the three
Clebsch potentials \cite{lamb,rasetti,bistro}, $\varphi$, $\Psi$ and
$\Psi^*$. Notice that, although the fluid is described in terms
of the 4-component field $(\rho,\varphi, \Psi, \Psi^*)$, there are no
fundamental `charges' and all dynamical effects derive from the possible
states of motion of the fluid.

In quantum hydrodynamics
$(\rho,\varphi)$ and $(\Psi, \Psi^*)$ are pairs of canonically conjugated
variables, i.e.
\BE
[\rho({\bf{r}}),\varphi({\bf{r}}')]= i\delta^{3}({\bf{r}}-{\bf{r}}')
\EE
and
\BE
[\Psi({\bf{r}}),\Psi^*({\bf{r}}')]= 
\delta^{3}({\bf{r}}-{\bf{r}}')
\EE
all other commutators being zero. In this way, one obtains
Landau's relations for 
the commutators of the velocity components
\BE
[{\bf{u}}_x({\bf{r}}),{\bf{u}}_y({\bf{r}}')]= {{1}\over{i\rho}}
\delta^{3}({\bf{r}}-{\bf{r}}') {\bf{\zeta}}_z
\EE
where 
\BE
\zeta=\nabla{\rm x} {\bf{u}}
\EE
is the vorticity.

In the incompressibility limit, where $\rho=\bar{\rho}$ 
and the phase $\varphi$ are constant
throughout the volume
of the fluid, 
 the fluid Hamiltonian density reduces to its `roton' part
\BE
H_{\rm roton}= 
-{{1}\over{8\bar{\rho} }}(\Psi^* \nabla \Psi - \Psi \nabla \Psi^*)^2
\EE
In a cubic box of volume $\vol$, one can expand in plane waves 
\BE
\Psi=  {{1}\over{\sqrt {\vol} }} \sum_{\bf k} b_{ {\bf{k}} }
e^{+i{\bf{k}}\cdot {\bf{r}} }~~~~~~~~~~~~
\Psi^*=  {{1}\over{\sqrt {\vol} }} \sum_{\bf k} b^*_{ {\bf{k}} }
e^{-i{\bf{k}}\cdot {\bf{r}} } 
\EE
with 
\BE
         [b_{ {\bf{k}} },b^*_{ {\bf{l}} }] = 
\delta_ { {\bf{k}} {\bf{l}} }
\EE
so that
by integrating the Hamiltonian density 
over the whole volume one gets quadratic 
         $b^*_{ {\bf{k}} }b_{ {\bf{k}} }$ 
terms. These can be used to define
a free-roton Hamiltonian
\BE
{\cal H}^{(o)}_{\rm roton}= 
{{1}\over{8\bar{\rho} \vol}}
\sum_{ {\bf {k}}{\bf {l}} } 
({\bf{k}}^2 + {\bf{l}}^2) 
         b^*_{ {\bf{k}} } b_{ {\bf{k}} }
\EE
that, after converting the discrete sum into an integral, gives finally
\BE
\label{final}
{E}_{\rm roton}({\bf{k}})= 
{{1}\over{8\bar{\rho}}} \int {{d^3 {\bf{q}} }\over{(2\pi)^3 }}
 ({\bf{k}}^2 + {\bf{q}}^2) 
\EE
Eq.(\ref{final}) is formally divergent, so that, 
to extract the relevant 
values of $\Delta$ and $\mu$ in Eq.(\ref{roton}), 
one has to find out a 
suitable cutoff momentum for the single-roton excitations, say
$|{\bf{q}}|=q_{\rm max}$. 
If this is done, we obtain
\BE
\label{ddelta}
        \Delta= {{q^5_{\rm max} }\over{ 80 \pi^2 \bar{\rho} }}
\EE
and 
\BE
\label{mumu}
         \mu= {{24 \pi^2 \bar{\rho} }\over{ q^3_{\rm max} }}
\EE
As a consequence of the introduction of $q_{\rm max}$, 
the uncertainty associated with any cutoff procedure will only allow
an order of magnitude estimate of $\Delta$ and $\mu$ and of the regime 
of parameters associated with the 
Lorentz-covariant condition $\Delta=\mu$ \cite{bethe}.

To obtain an estimate of $q_{\rm max}$, we observe that
using the  operators $b^*_{ {\bf{k}} }$ and $b_{ {\bf{k}} }$ one can 
construct the roton Fock space labelled by $N_{ {\bf{k}} } $, the eigenvalue of 
the roton number operator $b^*_{ {\bf{k}} } b_{ {\bf{k}} }$, the single-roton
states corresponding to 
$N_{{\bf{k}}}=1$. Now, by computing the value of
the vorticity vector in a single-roton state, one gets the idea of the roton "...
as a steady rotational
motion of the fluid, capable of moving as a `vortex' through the liquid"
\cite{ziman}. 

This observation arises by inspection of the single-roton
wave functions expressed in cylindrical co-ordinates $(r,\theta,z)$. In this
case, by introducing the squared transverse momentum
$\kappa^2=k^2_x + k^2_y$, the azimuthal quantum number $\nu$, 
and the z-component of the momentum $k_z$, one finds the equivalent form
for the free-roton Hamiltonian
\BE
{\cal H}^{(o)}_{\rm roton}= 
{{1}\over{8\bar{\rho} \vol}}
\sum_{ \kappa,\kappa',\nu,\nu',k_z,k_z' } 
(\kappa^2 + k^2_z + (\kappa')^2 + (k'_z)^2) 
         b^*_{ \kappa \nu k_z } b_{ \kappa \nu k_z }
\EE
whose single-particle eigenvalues 
\BE
         E_{\kappa \nu k_z}= \Delta + {{\kappa^2 + k^2_z}\over{2\mu}}
\EE
have the same $\Delta$ and $\mu$ as 
in Eqs.(\ref{ddelta}) and (\ref{mumu}) with
$q^2_{\rm max}=(\kappa^2 + k^2_z)_{\rm max}$.
 This can be understood by first
noticing that the quantities $\Delta$ and $\mu$ can be
expressed in terms of the number of quantum states with
$|{\bf{q}}|\leq q_{\rm max}$, say $g(q_{\rm max})$, as
\BE
\Delta \sim q^2_{\rm max} {{g(q_{\rm max})}\over{\bar{\rho} \vol}}
\EE
and
\BE
\mu \sim {{\bar{\rho} \vol}\over{  g(q_{\rm max})}}
\EE
Further, using the results of ref.\cite{patria}, one finds the same 
leading behaviour 
\BE
{{g(q_{\rm max})}\over{\vol}}\sim {{q^3_{\rm max}}\over{6\pi}}
\EE
by switching from cubical to cylindrical geometry.
Finally, averaging over all possible orientations gives
$(k^2_x)_{\rm max}=(k^2_y)_{\rm max}=(k^2_z)_{\rm max}$ and one obtains
$q^2_{\rm max}\sim {{3}\over{2}}(\kappa^2)_{\rm max}$.

In cylindrical coordinates, 
the single-particle wave functions can be expressed as 
\BE
\psi=\psi_{\kappa,\nu,k_z}(r,\theta,z)= {\cal N}
J_{\nu}(\kappa r)e^{i\nu \theta} e^{i k_z z}
\EE
where
$J_{\nu}(\kappa r)$ are Bessel functions and ${\cal N}$ a normalization
factor. In this case, one finds $u_r=u_z=0$ with
the only non vanishing component of the velocity being
\BE
u_\theta={{1}\over{2ir}}(
\psi^*{{\partial \psi }\over{\partial \theta}} -
\psi{{\partial \psi^* }\over{\partial \theta}} )= {\cal N}^2 
{{\nu J^2_\nu(\kappa r)}\over{r}}
\EE
Analogously, $\zeta_r=\zeta_\theta=0$ and 
the only non-zero component of the vorticity is
\BE
\zeta_z= {{1}\over{r}} {{\partial(r u_\theta) }\over{\partial r}}= 
{\cal N}^2 {{\nu}\over{r}} {{d}\over{dr}} J^2_\nu(\kappa r)
\EE
By introducing the vortex radius $R$ as the value at
which $u_\theta(r=R)=0$, and using the relation 
\BE
\int^{R}_{0} r dr J^2_\nu(\kappa r)= {{R^2}\over{2}} J^2_{\nu +1}(\kappa R)
\EE
(with $J_\nu (\kappa R)=0$) we can set finally 
${\cal N}=
\sqrt{ {{2}\over{R}} } 
{{1 }\over{ J_{\nu +1}(\kappa R) }}$. 

Therefore, a reasonable value of the cutoff momentum 
for single-roton states is
\BE
q_{\rm max} \sim \kappa_{\rm max} \sim {{1}\over{R_{\rm min} }}
\EE
where $R_{\rm min}$ denotes the minimum
transverse size of the thinnest vortices that can be established 
in the superfluid. These are the so called `vortex filaments', whose
transverse size can be obtained from  ref.\cite{landau2} in terms of
the particle density $n$ and of the scattering length $a$. 
 In this case, for a dilute
(`almost ideal') Bose condensate, the filament `core' is 
\BE
                R_{\rm min}\equiv r_{\rm core} \sim {{1}\over{ \sqrt {na} }}
\EE 
so that, for $n={\bar n}$, we find
\BE
\label{qmax}
          q_{\rm max} \sim  {{1}\over{r_{\rm core} }} \sim \sqrt {\bar{n}a} 
\EE 
In this way, assuming for the mass of the fluid constituents the same
Eq.(\ref{mrmfp}) established for the quanta of the scalar condensate, one finds
\BE
\label{rhobar}
\bar{\rho}= m_\Phi \bar{n} \sim \bar{n}^2 a^2 
\EE
so that replacing relations (\ref{qmax}) and (\ref{rhobar})  
into Eqs.(\ref{ddelta}) and (\ref{mumu}), 
one gets the order of magnitude estimate
\BE
                    \Delta \sim \mu \sim \sqrt{\bar{n} a}
\EE
Therefore, by comparing with Eq.(\ref{mh}), 
we have identified
the relativistic regime $\Delta=\mu=M_H$
foreseen in the Introduction. In fact, 
$M_H$ coincides with the energy-gap $\Delta$ for 
vortex formation in a superfluid medium possessing the same density and the
same type of constituents as the scalar condensate. 

More generally, 
it is interesting to compare the possible regimes of a Bose superfluid,
made up of particles with mass $m$, 
in terms of the two dimensionless parameters 
\BE
\label{xx}
                 x= {{m}\over{n a^2}}
\EE
and 
\BE
                 \epsilon= \sqrt {n a^3} 
\EE
In terms of $x$, using Eq.(\ref{qmax}) for arbitrary $n$, 
Eqs.(\ref{ddelta}) and (\ref{mumu}) give
\BE
            \Delta \sim {{\sqrt{na} }\over{x}}
\EE
and 
\BE
\label{x2}
             \mu \sim x \sqrt{na} \sim x^2 \Delta 
\EE
so that, as anticipated, the
Lorentz-covariant condition $\Delta\sim \mu$ corresponds to $x\sim 1$. 

Notice that the momentum 
$\delta\sim 1/r_{\rm mfp}\sim {n}a^2$, related to the transition
from the phonon branch to the roton-like
excitations of Eq.(\ref{roton}) can be used to obtain
the sound velocity from the relation
\BE
c_s \delta \sim \Delta 
\EE
so that (in units of $c$) 
\BE
\label{sound}
    c_s \sim 
{{1}\over{x \epsilon}} 
\EE
Using Eq.(\ref{x2}), the non-relativistic limit, where
$\Delta \ll \mu$, is seen to correspond to very large values of $x$ such 
that also $1 \ll x\epsilon$. In this case, the sound velocity
becomes
\BE
    c_s \sim {{1}\over{x \epsilon}} 
\sim {{ \sqrt{ {n}a} } \over{m}} \ll 1
\EE
which is the Lee-Yang-Huang result for a dilute hard-sphere Bose gas 
\cite{lhy}.
 On the other hand, a Lorentz-covariant form of the massive branch
requires a value $x \sim 1$ in Eq.(\ref{x2}). This, when replaced in 
Eq.(\ref{sound}), produces the anticipated
 vastly superluminal sound velocity Eq.(\ref{cs})
\BE
c_s \sim 1/\epsilon 
\EE
that diverges in the limit where $\epsilon \to 0$ and the scale 
$\Delta \sim \mu \sim \sqrt{na}=$ is kept fixed.

The above relations can also be compared with superfluid $^4$He. 
Although this is {\it not} a dilute system
(with typical values 
$a \sim 2.7\cdot 10^{-8}$ cm and $n\sim 10^{23}$ cm$^{-3}$, one gets 
$\epsilon\sim 1.4$), we find,  
nevertheless, a good agreement with our picture. 
In fact, using the experimental values 
\cite{tilley}
$(\mu)_{\rm exp}= 0.16 m_{\rm He}$ and 
$(\Delta)_{\rm exp}= 7.4\cdot 10^{-4}$ eV in Eq.(\ref{x2}) 
one can obtain an experimental value of $x$, say 
\BE
\label{exp}
            x_{\rm exp}\equiv \sqrt {  
{{(\mu)_{\rm exp}c^2}\over{(\Delta)_{\rm exp} }} } \sim 10^6
\EE
that, if used in Eq.(\ref{sound}), 
produces a value
$c_s \sim 0.7\cdot 10^{-6}$ 
in good agreement with the experimental
result for the sound velocity 
 $(c_s)_{\rm exp}=239$ m/sec \cite{tilley}. Finally, the theoretical input
prediction from Eq.(\ref{xx}) 
\BE
x_{\rm th}\equiv {{m_{\rm He} }\over{na^2}} \sim 3\cdot 10^{6}
\EE
is also in fairly good agreement with the experimental result Eq.(\ref{exp}), 
thus confirming the overall consistency of our picture.

\section{Summary and outlook}

Taking into account
the two-valued nature \cite{legendre,pmu} of the zero-4-momentum
connected propagator in the broken phase, 
 one gets the idea of a true
physical medium that contains,
for ${\bf{k}}\to 0$,
two types of excitations: a massive one, 
whose energy $E_a({\bf{k}}) \to M_H$  and that 
corresponds to the usual Higgs boson field,
and a gap-less one whose energy
$E_b({\bf{k}}) \to 0$. 
 The overall picture is similar to the 
coexistence of phonons and rotons in superfluid $^4$He that, in fact,
 is usually
considered the condensed-matter analogue of the Higgs condensate. 

Now, the gap-less branch is naturally interpreted in terms
of the collective density
fluctuations of the system \cite{weak,hierarchy,drift}.
These dominate the physical spectrum 
for ${\bf{k}}\to 0$ and their
wavelengths are larger than $r_{\rm mfp}$, 
the mean free path for the condensed quanta. 

On the other hand, 
the continuum limit of quantum field theory corresponds to the ideal case
of an incompressible fluid so that
the massive energy spectrum 
$E_a({\bf{k}})$ extends down to ${\bf{k}}=0$. 

In this paper, 
following the original Ziman's \cite{ziman} approach and using the formalism of 
ref.\cite{mech}, we have proposed the interpretation 
of the massive branch as a roton, i.e. as an elementary excitation that, 
differently from the phonons associated with the irrotational motions of the
superfluid, arises in connection with a non-zero (`bulk')
vorticity. This interpretation requires 
the energy spectrum Eq.(\ref{roton}) to exhibit values of $\Delta$ and
$\mu$ such that
\BE
\Delta\sim \mu \sim M_H \sim \sqrt { \bar{n} a}
\EE
In turn, this relation depends on the 
peculiar condition  Eq.(\ref{mrmfp})
\BE
\label{11}
m_\Phi \sim {\bar n}a^2
\EE
 between the mass $m_\Phi$ of
the elementary condensing
phions, their equilibrium number density $\bar{n}$ and their
scattering length $a$. 
Relation (\ref{11}) implies that the phion mean free path 
$r_{\rm mfp} \sim 1/(\bar{n}a^2)$ is
of the same order as the phion Compton wavelength $1/m_\Phi$ and
is naturally found in the weakly first-order scenario of ref.\cite{mech}
where the broken phase is represented as a dilute Bose condensate for which
$\epsilon=\sqrt{{\bar n}a^3} \ll 1$, the continuum limit corresponding to 
$\epsilon \to 0$ with ${\bar n}a=$ fixed.

Of course, being used to consider Lorentz covariance 
an exact built-in requirement, 
an energy spectrum as in Eq.(\ref{roton}) with
$\Delta=\mu$ may seem more or less
trivial. For instance, starting from the original annihilation and creation 
operators for the elementary quanta of the unphysical empty vacuum state
$|o\rangle$, it is easily discovered within the standard
covariant generalization of the Bogolubov method \cite{mech}.  
However, our analysis shows that there is an alternative viewpoint. 
In fact, within quantum hydrodynamics the same result is far from 
being trivial and is only 
recovered in the special case (\ref{11}). In this modified perspective, 
a Lorentz-covariant 
massive spectrum corresponds (with some approximations and in a
certain range of wavelengths) 
to vortex formation in a
superfluid medium possessing a well defined pattern of scales. 

We conclude by observing that the proposed identification of the massive 
Higgs boson as a roton and of $M_H$ with the
energy-gap for vortex formation in the superfluid vacuum is not a mere
exercise. In fact, in a quantum-hydrodynamical description of the scalar
condensate based on the hierarchical structure of scales Eq.(\ref{scales}),
the single-roton states have a natural cutoff momentum at 
\BE
   q_{\rm max} \sim \sqrt{ \bar{n} a} \sim M_H \ll \Lambda
\EE 
(with $\Lambda\equiv 1/a$). 

Therefore, accepting our interpretation, the Higgs boson momentum would be 
physically cut off at values that are infinitely smaller than $\Lambda$ 
(by a factor $\sqrt{ \bar{n} a^3} \to 0$) when approaching the continuum limit.
This implies, for instance, that any self-energy part that grows 
quadratically with the phase space, say
\BE
     \Pi_H= \sum_n c_n \lambda^n q^2_{\rm max}~,
\EE
is much smaller than $M^2_H$, for a weak self-coupling $\lambda$. Thus
$M_H$ emerges as a naturally intermediate, fixed energy scale associated with
the condensed vacuum and, in this sense, by 
treating the Higgs condensate as a real physical medium, one can find a 
solution of
the so called {\it naturalness} problem (i.e. why $M_H$ is so much smaller than 
$\Lambda$) without any artificial fine-tuning
of the basic parameters. Such fine-tuning problems, instead, appear
in the standard approach where the massive Higgs boson is regarded as
an ordinary elementary particle propagating in the vacuum and its maximum
momentum is identified with the ultimate ultraviolet cutoff
$\Lambda$ of the theory.


\begin{thebibliography} {99}
\bibitem{physical}
By `physical' we mean that the normalization of the vacuum field
$v_R$ is such that 
the quadratic shape of $V_{\rm eff}(\phi_R)$ at $\phi_R=v_R$ is precisely 
given by the physical Higgs mass $M^2_H$.
\bibitem{bethe}
An effective cutoff in momentum space 
has been widely used in the literature before
the introduction of the idea of renormalization. For instance Bethe's 
\cite{bethe2} and Welton's \cite{welton} treatments of Lamb's shift, both
based on the choice $q_{\rm max}=m_e c$, 
capture the essential physics and the exact order of magnitude of the effect
although it is hard to make them into a fully satisfactory calculation.
\end{thebibliography}
\end{document}